\documentstyle[11pt,aaspp,psfig,fullpage]{article}
\begin{document}
\title{Synchrotron Radiation as the Source of Gamma-Ray Burst Spectra.}

\author{Nicole M. Lloyd \& Vah\'e Petrosian}
\affil{Center for Space Science and Astrophysics, Stanford University,
Stanford, California 94305}

\begin{abstract}
We investigate synchrotron emission models as the source
of gamma-ray burst spectra.  We show that including
the possibility for synchrotron self
absorption,
a ``smooth cutoff'' to the electron energy distribution,
and an anisotropic distribution for the electron pitch angles
produces a whole range of low energy spectral behavior.
 In addition, we show that the procedure of spectral
 fitting to GRB data over a finite bandwidth can introduce
 a spurious correlation between spectral parameters - in particular, 
the value of the peak of the $\nu F_{\nu}$ spectrum, $E_{p}$, and the
low energy photon spectral index $\alpha$ (the lower $E_{p}$ is,
the lower (softer) the fitted value of $\alpha$ will be).
From this correlation and
knowledge of the $E_{p}$ distribution, we show how
to derive the expected distribution of $\alpha$.  We
show that optically thin synchrotron models with an
isotropic electron pitch angle distribution can
explain the distribution of $\alpha$ below $\alpha=-2/3$.
This agreement is achieved if we relax the unrealistic assumption of 
the presence of a sharp low energy cutoff in the
spectrum of accelerated
electrons, and allow for a more gradual break.
We show that this low energy portion of the electron spectrum can
be at most flat.
We also show that optically thin synchrotron models
with an {\em anisotropic} electron pitch angle
distribution can explain all bursts with $-2/3 \la \alpha \la 0$.
The very few bursts with low energy spectral indices that fall
above $\alpha=0$ may be due the presence of a
the synchrotron self-absorption frequency entering the lower end of
the BATSE window.  Our results also predict a particular
relationship between $\alpha$ and $E_{p}$ during
the temporal evolution of a GRB.  We give examples of 
spectral evolution in GRBs and discuss how the behavior 
are consistent with
the above models.
\end{abstract}

\section{Introduction}
The spectrum of a gamma-ray burst (hereafter, GRB) is a crucial element
 in understanding
the nature of the event.
  It can provide information about
the burst energetics, the local magnetic field, particle distributions and
acceleration mechanisms, and the overall expansion of the fireball.
Unlike the extreme variation in the light curves, the spectra
of GRBs are fairly homogeneous.
 The photon spectrum of most GRBs in the BATSE spectral
 energy range can be parameterized
by a broken power law with 
low and high energy photon spectral indices, $\alpha$ and $\beta$,
 and a break energy, $E_{b}$ (e.g. Band et al., 1993). 
For the majority of bursts, $\alpha > -2$ and $\beta < -2$, in
which case the $\nu F_{\nu}$ spectrum has a maximum
at $E_{p}$ that represents where
the burst emits most of its energy; for a Band
spectrum, $E_{p} = (2+\alpha)E_{b}/(\alpha-\beta)$.
 In most emission models, $E_{p}$ (or $E_{b}$)   
usually reflects some characteristic {\em electron} energy. 
In models of GRB emission with a power
 law energy spectrum for the radiating particles, it
could signify the presence of 
  a cutoff in the underlying particle distribution, or the energy
  above which all electrons ``cool'' - rapidly radiating all of their 
  kinetic energy within the source (see Piran, 1999, for a review).
 The high energy photon index, $\beta$, usually
 reflects the steepness
of the particle
energy distribution.
The value of the low energy photon index, $\alpha$, on the
other hand varies significantly from model to model and can be
due to several factors.  Therefore, it has 
greater potential to distinguish between the
different scenarios for GRB emission.  
However, despite years of availability of high quality
spectral data from GRBs,  the burst emission process remains
ambiguous.

It was suggested (e.g. Katz, 1994) that {synchrotron emission} is
a likely source of radiation from GRBs, and later shown (Tavani, 1996)
that an {\em optically thin} synchrotron spectrum 
from a power law distribution of relativistic
electrons with a {\em sharp minimum energy cutoff}  ($N(\gamma) \propto
\gamma^{-p}, \gamma > \gamma_{m}$, where
$\gamma$ is the electron Lorentz factor) and with an {\em isotropic pitch
angle distribution}
provides a good fit to some bursts.
However, some features seen in the low
energy portion of GRB spectra can not be explained
by this simple synchrotron model.
  This model predicts that $E_{p}$ is related to
  the minimum value of the electron Lorentz factor, $\gamma_{m}$
  ($E_{p} \propto 
  \gamma_{m}^{2}h\nu_{B}$, where $\nu_B=eB/2\pi m_{e}c$ is the gyrofrequency in a
  magnetic
field with mean perpendicular component
 $B$), so that the asymptotic
value of the low energy
photon index, $\alpha$,  should be a constant value of $-2/3$ (e.g., Pacholczyk, 1970). 
 Hence, if
our spectral fits are determining
the asymptote accurately, the
$\alpha$ distribution should be a very narrow distribution centered
about $-2/3$. 
\begin{figure}[t]
\leavevmode
\centerline{
\psfig{file=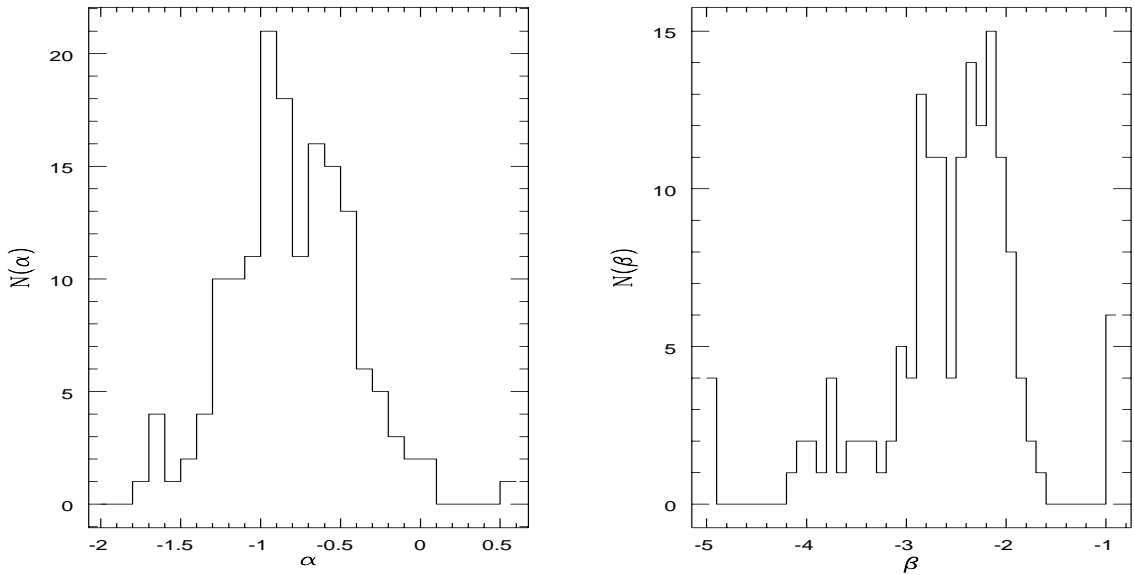,width=1.\textwidth,height=0.5\textwidth,angle=270}}
\caption{Distributions of the low and high energy photon spectral
indices $\alpha$ (left panel, $N(\alpha)$) and 
$\beta$ (right panel, $N(\beta)$), from Preece et al.
(1999).}
\end{figure}
  Figure 1 shows the time averaged distributions of $\alpha$ and 
  $\beta$ taken from Preece et al. (1999).  
 The $\alpha$ distribution clearly appears to disagree with this
 simple synchrotron 
model.
It has also been claimed (e.g. Crider
 et al., 1997) that spectral evolution of $\alpha$ (and $E_{p}$, for that matter)
  throughout a burst is inconsistent with the standard synchrotron scenario, at least
in the context of a single (external) shock model. 
{Consequentially, other models - usually involving Compton scattering
(Brainerd, 1994, Liang \& Kargatis, 1996) -
 are invoked to explain this ``anomolous''
spectral behavior.}
Another difficulty with the synchrotron model stems from theoretical
considerations.  For typical spectral peak energies $E_{p} \sim$ few$\times 100$ keV,
and ``typical'' values for the electron Lorentz factors $\gamma \sim 100$ and bulk
Lorentz factor $\Gamma \sim 100$, magnetic fields can reach values of about
$10^{7} G$ (such fields are also reached simply assuming
equipartition between photon, electron and magnetic field
energy density).  As a result, the synchrotron lifetime in
the observer's frame ($\tau_{s} = (\gamma/\dot{\gamma_{s}})(1+z)/\Gamma) 
\la 10^{-6} s$, where $\dot{\gamma_{s}}\propto \gamma^{2}B^{2}$ is
the synchrotron energy loss rate, $\Gamma$ is the
bulk Lorentz factor of the medium, and $z$ is the
GRB redshift) is much shorter than the physical timescales relevant in
the standard fireball model for GRBs (again, see Piran, 1999 for a review),
as well as the detector integration time.
  Thus, one expects these
radiative loss effects to be apparent in the GRB spectrum.
For example,  
if the electrons are injected in a source
region where those with energies $\gamma > \gamma_{\rm c}$ lose all their
energy to radiation (while those with $\gamma<\gamma_{\rm c}$ escape
the emission region before suffering significant losses of their kinetic
energy), then one obtains the so-called {\it cooling} 
 spectrum which has an additional break at a photon energy 
$E_{c} \propto \gamma_{\rm c}^2h\nu_B$.  Depending on the relative values of 
$\gamma_{m}$ and $\gamma_{c}$,
one has a variety of values for the spectral indices.  For example, if
$\gamma_{\rm c} > \gamma_{\rm min}$ (and we have the same power law distribution
of electron energy mentioned above) then
one has an index $\beta'= -(p + 2)/2$ above $E_{\rm c}$.
  In the case when $\gamma_{\rm c} <
\gamma_{\rm min}$, {\em all} of the injected electrons
lose their energy to radiation and we therefore have
an index $\alpha' = -3/2$ between 
$\gamma_{\rm c}$ and
$\gamma_{\rm min}$.  The peak of the
$\nu F_{\nu}$ spectrum, $E_{\rm p}$, could represent either the cooling or the
minimum Lorentz factor break, depending on their relative values and their
positions in the BATSE spectral window (see, e.g. , Sari, Piran \& Narayan
1998).  Hence, in a cooling spectrum, we expect either $\alpha$ or
$\beta$ to be about -1.5, which is not borne out
by the observations as seen in Figure 1.  Ghissilini et al. (1999) use
this argument to rule out synchrotron emission in favor of a quasi-thermal
inverse Compton model.  We agree that the simple cooling spectrum
is ruled out by the observations.  However, we also believe that this
does not rule out synchrotron emission in general,
and that all of these above mentioned discrepancies arise
from incorrect simplifying assumptions in the models.

Let us first consider the arguments based on the discrepancy
with the cooling spectrum.  We believe the reason for this discrepancy
 lies in the
artificial separation between particle acceleration and radiation.
In this model
 particles are accelerated in one region, and then either the
 acceleration processes automatically cease or the particles are
 injected into another region in which they radiate away all of
 their energy.
We
  believe that such a  model of the particle
  acceleration and radiation processes is an unrealistic situation. 
It is 
  likely that both acceleration and cooling or radiation
  processes take place behind the 
  shock 
continually throughout the emission episode. The acceleration rate
$R_{acc}$ and the 
synchrotron loss rate $R_{syn} = \gamma/\dot{\gamma_{s}}$, where
$\dot{\gamma_{s}}$ is defined in the previous paragraph,
compete with each other in forming the instantaneous 
spectrum of the electrons which produces the observed synchrotron spectrum. 
Since the synchrotron loss  increases rapidly with energy, its effect is to 
inhibit acceleration beyond some maximum value $\gamma_{\rm max}$ where the 
two rates become equal, $R_{acc}(\gamma_{\rm max}) =
R_{syn}(\gamma_{\rm max})$. Thus, we expect a power law electron 
spectrum for $\gamma<\gamma_{\rm max}$, with 
a relatively sharp cutoff outside $\gamma_{\rm max}$. 
In reality, the spectrum of the accelerated electrons could be more
complex and may lead to the formation of a plateau just below
$\gamma_{max}$ or to other features (e.g., 
see Petrosian \& Donaghy, 2000). 
Considering 
observations up to the GeV range of some GRBs by the EGRET instrument on CGRO, 
$\gamma_{\rm max}$ must be large enough so that the energy of the corresponding 
synchrotron 
photons will far exceed the BATSE range. Thus in this model $E_{p}$ is related to 
$\gamma_{m}$, the minimum energy of the accelerated electrons,
 and not to any cooling break.

We point out that the high energy photon indices for a
simple cooling spectrum and the so-called ``instantaneous'' spectrum 
(in which cooling effects are negligible) are
$-(p+2)/2$ and $-(p+1)/2$, respectively.  Observed values
of $\beta \sim -2.1$ indicate values of $p \sim 2.2$ 
for the simple cooling model and $3.2$ for the  
instantaneous 
spectrum.  Recent studies of particle acceleration via
the Fermi mechanism in
relativistic shocks determine a ``universal'' index of $p \sim 2.2$ 
(e.g., Gallant 
  et al., 2000, Guthman et al., 2000), which is more consistent
the values expected from the cooling model, and this has been
used as an argument against the instantaneous spectrum.  However,
we note that $\beta$ is usually not very well constrained by
the data and the $1\sigma$ errors are frequently of the order $\pm 1/2$, so that
observations cannot distinguish between the cooling and instantaneous spectra.
Moreover, the cooling spectrum itself is not consistent with
the universal index for the significant fraction of bursts with
$\beta \la -2.5$.  In any case, this universal index
is based on a very simplified model;
particle acceleration in the context of
 internal shocks has not been carefully studied, and it is unclear whether
 the Fermi mechanism will even work at all (see, e.g. Lyutikov \& Blackman, 2000
 and  Kirk et al., 2000, the latter of whom show that including a magnetic field
 at the acceleration site will steepen this universal index).
 In most of our analysis below, we will use the
 instantaneous  synchrotron spectrum, and not 
the simple cooling spectrum; we will
show that this instantaneous synchrotron spectrum
can indeed explain the behavior seen in the BATSE spectral data.

  Another simplification in standard synchrotron models
  that leads to discrepancies between the theory and the observations
  is the assumption
  that the GRB emission region is optically thin.  Under certain
  conditions, the medium may in fact
  become opaque, e.g.
  to synchrotron self-absorption (discussed in \S 2.1).  This will produce a steep low
 energy cutoff in the photon spectrum.  This is one possible
 explanation for those bursts with $\alpha$ values above
 the so-called 
 ``line of death'', $\alpha>-2/3$ (Preece et al., 1998a)
 
  A third unrealistic simplification 
  involves the electron spectrum at low energies.  The asymptote
  of the optically thin photon spectrum $\alpha=-2/3$ is obtained if one
  assumes that the electron spectrum cuts off sharply below $\gamma_{m}$.
It is unlikely that the acceleration process will give rise to such a 
spectrum, which is subject to plasma instabilities.  Particles in front
of a relativistic shock - after crossing it - will acquire a Lorentz
factor comparable to the bulk Lorentz factor of the shock.  Subsequent
interaction with the turbulent medium will accelerate them to higher
energies (up to $\gamma_{max}$ described above), but the same
processes will decelerate some to lower energies (see, e.g., Park
\& Petrosian, 1995, and references cited therein).  Plasma instabilities
(e.g. the two stream instability) can further smooth this peaked
spectrum (and in some cases
give rise to a nearly flat spectrum for $\gamma<\gamma_{m}$).
Inclusion of such effects will clearly modify the behavior of the low
energy synchrotron emission.

 A fourth assumption in the simple synchrotron model, which may not
 always be correct, is that the pitch angle distribution of the 
 electrons is isotropic.  An isotropic distribution is expected
 if the acceleration mechanism is more efficient in changing
 the pitch angle rather than the energy of the particles, which
 is commonly the case.  However, in low density  high magnetic
 field plasmas, the opposite may be true so that a  highly anisotropic
 distribution of pitch angles is expected; as a result, the low
 energy synchrotron spectrum
 will differ from the usual simple model.  Note that 
 as long as the magnetic field lines have random orientations,
  the total emission in the rest frame will be 
 isotropic.
 For details, see Epstein (1973). 
 
  In what follows, we will show that
 incorporating all of these effects can lead to a wide variety of
 low energy GRB spectral behavior. 
  However, there is another important cause of the discrepancy
between the predicted and observed spectral behavior, which arises as a result
of the finite bandwidth of the
instrument and the spectral fitting procedure.  One cannot assume that the
spectral fits to phenomenological models are able to
accurately determine the asymptotic values of the spectral indices.
This is because the actual spectra do not show sharp breaks
at $E_{p}$ (or $E_{b}$) from the high energy spectral index to the low energy spectral
index - rather, there is a smooth transition between the two indices.
The fitted values of the
low energy (and high energy, for that matter) spectral index depends
on how far the spectral window extends below (or above) $E_{p}$.
We must take these effects into consideration when testing any model.

In this paper we show that when these procedural effects as well as
more realistic synchrotron models are incorporated, synchrotron radiation
can accomodate both
the shape of the distributions and
temporal evolution of GRB spectral parameters, and the discrepancies
discussed above are resolved. Of course this does not {\em prove} the synchrotron model
is the unique emission mechanism for GRBs, but merely shows that
it is consistent with the existing data.
In \S 2, we discuss the various spectral shapes obtained from a general form
for synchrotron emission, allowing for the possibility for self-absorption,
 a smooth (instead of sharp) cutoff to the electron energy distribution,
and a small pitch angle distribution. 
 In \S 3, we consider the instrumental and fitting effects and point out a  
correlation that exists between $\alpha$ and $E_{p}$ as determined by
the Band spectrum (Band, 1993).
 In \S 4, we use this relationship and knowledge
of the $E_{p}$ distribution to determine the expected $\alpha$ distribution,
and compare this with the observed distribution.
We also discuss the presence of an absorption cutoff and evidence
of small pitch angle scatering observable by BATSE
in some GRBs to explain particularly those bursts with $\alpha>-2/3$.
These results also predict a certain relationship between $\alpha$
and $E_{p}$ during the temporal evolution of a GRB.
In \S 5, we give examples of spectral evolution in GRBs and discuss whether
these are
consistent with synchrotron emission models.
A summary of our conclusions are given in \S 6.

\section{Synchrotron Spectra}

Synchrotron radiation will occur when relativistically
charged particles have a component of their velocity perpendicular
to a local magnetic field (i.e. a non-zero pitch
angle).  The importance of synchrotron emission
(compared to, say, inverse Compton or brehmstrahlung radiation) depends - 
among other things - on the strength of the magnetic field.
Given the relativistic nature of GRBs and probable physical conditions (e.g.
significant magnetic fields), it is likely that synchrotron radiation
plays a role in the emission from GRBs.
  The general form for an instantaneous synchrotron {\em energy} spectrum
from electrons with a (homogeneous)
 power law distribution of Lorentz factors with a {sharp
cutoff}, $N(\gamma) = N_{o}\gamma^{-p}\Theta(\gamma-\gamma_{m})$
(where $\Theta$ is the Heaviside step function), and with
an isotropic pitch angle distribution, is given by (e.g. Pacholczyk, 1970)
\begin{equation}
F_{\nu}= {\cal A} \nu^{5/2}
[\frac{I_{1}}{I_{2}}] \times  
[1.0 - {\rm exp}[-Q \nu^{-(p+4)/2}
I_{2}]]
\end{equation}
\begin{equation}
I_{1} = \int_{0}^{\nu/\nu_{m}}dx\ x^{(p-1)/2}
\int_{x}^{\infty} K_{5/3}(z) dz
\end{equation}
\begin{equation}
I_{2} = \int_{0}^{\nu/\nu_{m}}dx\ 
x^{p/2}
\int_{x}^{\infty} K_{5/3}(z) dz,
\end{equation}
\begin{figure}[t]
\leavevmode
\centerline{
\psfig{file=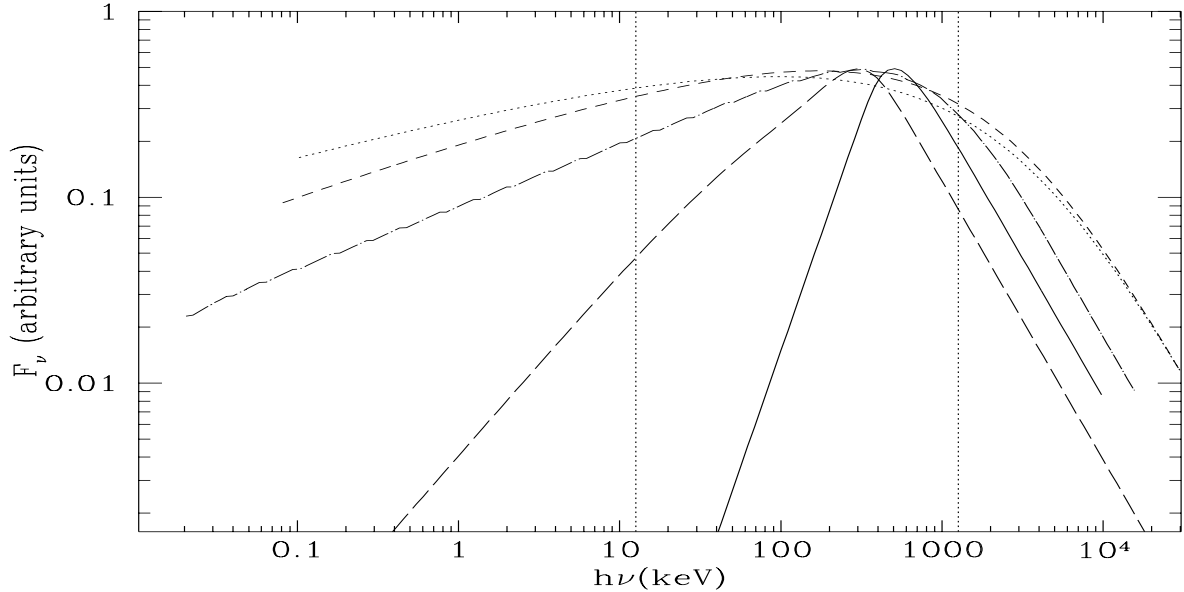,width=1.0\textwidth,height=0.5\textwidth,angle=270}}
\caption{Various synchrotron energy spectra, $F_{\nu}$ (in arbitrary
units), as a function
of energy $h\nu$ in keV.  The dot-dashed line is 
  the usual simple optically thin spectrum (with a sharp
  cutoff to the electron distribution, $q \rightarrow \infty$),
   while the dotted and short-dashed
  lines show optically thin spectra for  $q=0$ and $q=2$ cutoffs
  respectively; all of these are for an isotropic electron pitch
  angle distribution.  The solid line and long dashed lines show a self-absorbed
  spectrum
  for $\nu_{a}>\nu_{m}$ and the small pitch angle distribution
  case, respectively.  
The vertical lines mark the approximate width of the BATSE spectral window.
The spectra are normalized at their peaks to some
representative value and all have the same value
of the high energy spectral index, $p=4$.}
\end{figure}  
Here, 
we have assumed that the electrons are extremely relativistic, 
$\gamma_{m} \gg 1$, and that the magnetic
field at the source is randomly oriented, or that the emission is isotropic. 
The coefficient
${\cal A}$ is the normalization and contains factors of the magnetic field,
 $B$, bulk
Lorentz factor, $\Gamma$, and electron number,
$N_{o}$. The integrand $K_{5/3}(z)$ is a modified Bessel function of order
$5/3$.  The parameter
$\nu_{m}=((3/2)\Gamma \gamma_{m}^{2}\nu_{B})$, 
and  $Q$ is proportional to the optical
depth (to self-absorption) of the medium; for $\nu \gg \nu_{m}$, the
photon spectrum is self-absorbed at $\nu < \nu_{a} \propto
Q^{2/(p+4)}$.  The high energy optically
thin asymptotic behavior is the usual $F_{\nu}
 \propto \nu^{-(p-1)/2}$.
The low energy asymptotic forms of the function depend on the relative
values of $\nu_{m}$ and $\nu_{a}$:  
 $F_{\nu} \propto \nu^{5/2}$ for $\nu_{m} < \nu \ll \nu_{a}$, but
 is much flatter  - $F_{\nu} \propto \nu^{1/3}$ - for $\nu_{a} < \nu < \nu_{m}$.
 For very low frequencies, $\nu \ll {\rm min}[\nu_{a},\nu_{m}]$, 
 $F_{\nu} \propto \nu^{2}$.
We point out that the {\em photon} spectral index $\alpha$ is
$d$log($F_{\nu})/d$log$(\nu) - 1$.  
 Figure 2 shows the many different types of
 low energy behavior one can obtain from synchrotron
 emission.  The dot-dashed line is 
the usual simple optically thin spectrum, 
and the solid line shows a self-absorbed spectrum 
for $\nu_{a}>\nu_{m}$.  The dotted and short dashed lines
show optically thin spectra for a gradual (rather than sharp)
cutoff to the electron distribution, discussed in \S 2.2.
The long dashed line shows a spectrum in the case of small
pitch angle scattering, discussed in \S 2.3.
Since we are
  focusing
on the low energy spectral index $\alpha$, we have kept $\beta$ constant
and normalized the spectra at their peaks to a representative
value within the BATSE window approximated by the 
vertical dotted lines.

\subsection{Synchrotron Self Absorption}
Most treatment of synchrotron radiation from GRBs
has been in the optically thin case (see, however, Papathanassiou, 1998).
This is because synchrotron self-absorption 
requires ``extreme'' physical conditions in a GRB - particularly,
high magnetic fields and a high column density of electrons. 
For example, for $\nu_{a} < \nu_{m}$, the optical depth to
synchrotron self absorption is $\tau \sim (l/10^{13}cm) (n/10^{8}cm^{-3})
(B/10^{8}G)^{2/3}(\gamma_{m}/50)^{-8/3}(\Gamma/10^{3})^{3}(\nu_{obs}/10^{19}Hz)^{-5/3}$,
where $l$ and $n$ are the path length and particle density in the co-moving
frame, $\nu_{obs}$ is the absorption frequency in the observer's frame (note that
this frequency falls within BATSE's spectral window), and
we have assumed an electron energy distribution index $p=2$.
For a more detailed discussion of absorption frequencies in various
regimes, see Granot et al., 2000.
An example of a self-absorbed spectrum is
 given by the solid line if Figure 2.  
 As evident from the above expression, optical depths of order unity can
 be achieved within the BATSE spectral range with somewhat extreme values
 of the physical parameters, most notably the value of the magnetic field.
However, such high values of the magnetic field
 can be reached simply through
equipartition in internal shocks (see, e.g., Piran, 1999). 
  In general, we understand far too little about
 the generation of magnetic fields and the hydrodynamical conditions relevant
 for GRBs to rule out these extreme physical parameters.
 We also point out that the required values of $n$, $l$, and $\gamma_{m}$ lead to
 a Compton Y parameter of order unity or greater, thereby reducing the efficiency
 of synchrotron radiation in the observed range (Kumar, 1999, Piran, 1999).
 In this paper, we do not focus on
 how the conditions required for observable synchrotron self-absorption   
 are achieved, but rather investigate the consequences
 if they indeed are reached.  In any case, as shown in Figure 1 above
 and as we will see below, a self-absorbed spectrum may be applicable to
 only a small fraction of the GRB population.

 
 To test how well the optically thin and thick
 synchrotron spectra
 fit the data, we fit 11 bursts with 256 channel energy resolution
to the synchrotron spectral
form in equation (1).   In six cases, the {\em optically
thin} spectrum fit the bursts well.  But for five bursts, which
 have a low energy photon index $\alpha>-2/3$
(as determined
by fitting a Band spectrum, Preece et al., 1999), the
optically thin spectrum does not provide a good fit. 
For these bursts, we find that when
including a parameter characterizing the optical depth ($Q$
in equation 1), {\em the fits are improved significantly over
those to the optically thin spectra}.
Crider \& Liang, 1999,
also show self-absorption is consistent with GRB 970111, another
burst with a hard low energy spectral index ($\alpha>-2/3$). 
It should be noted, however, 
that in all of these bursts, $\nu_{m} > \nu_{a}$,
and $\nu_{a}$ was close to the edge of the BATSE window
(in fact, the presence of only a few bursts with $\alpha = 1$ or
$3/2$ (see Figure 1) indicates that in general $h\nu_{a} \la 50$ keV).  We discuss
the implications this has on the observed $\alpha$ distribution
below.  
\begin{figure}[t]
\leavevmode
\centerline{
\psfig{file=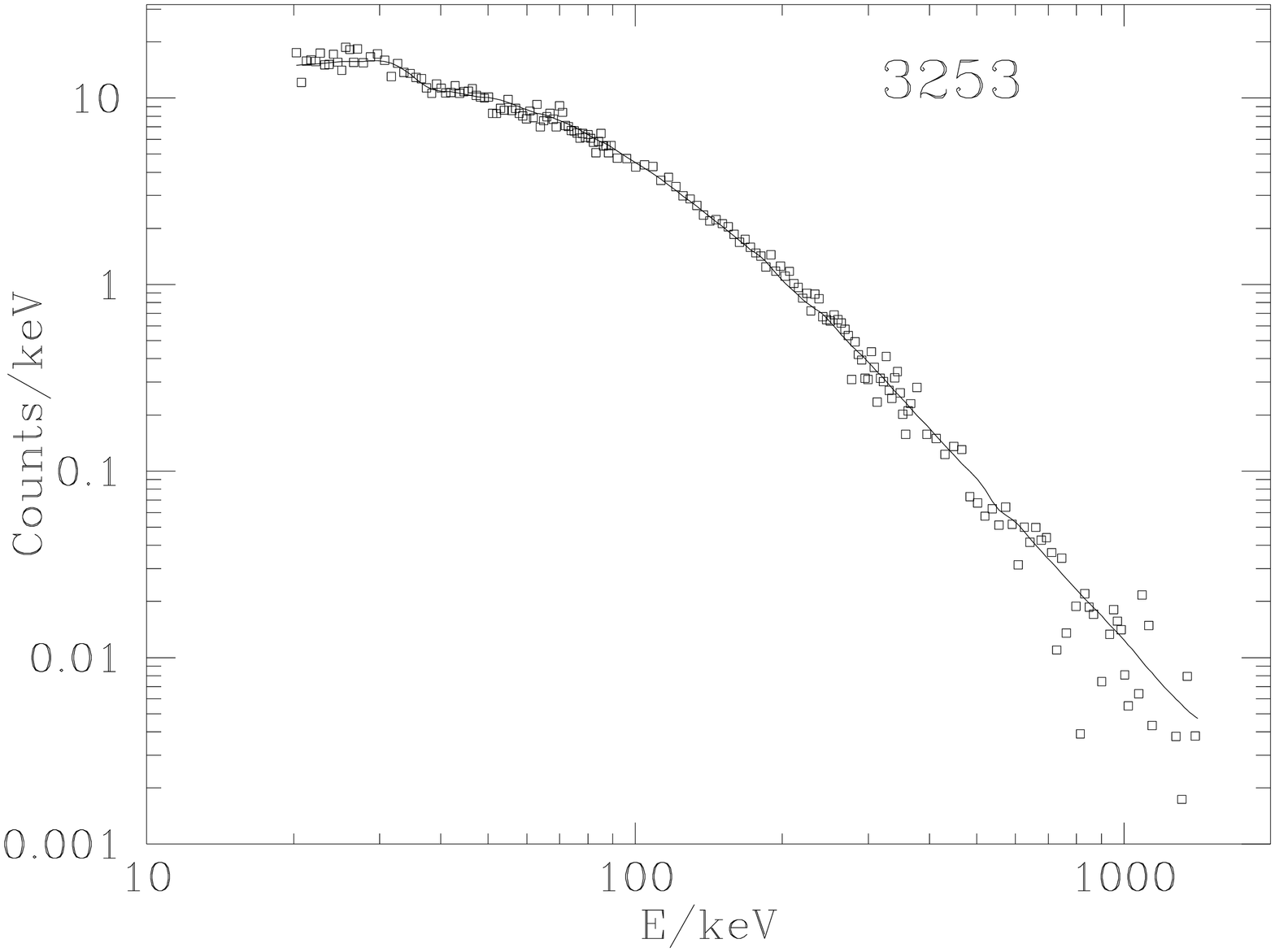,width=.5\textwidth,height=0.5\textwidth,angle=270}
\psfig{file=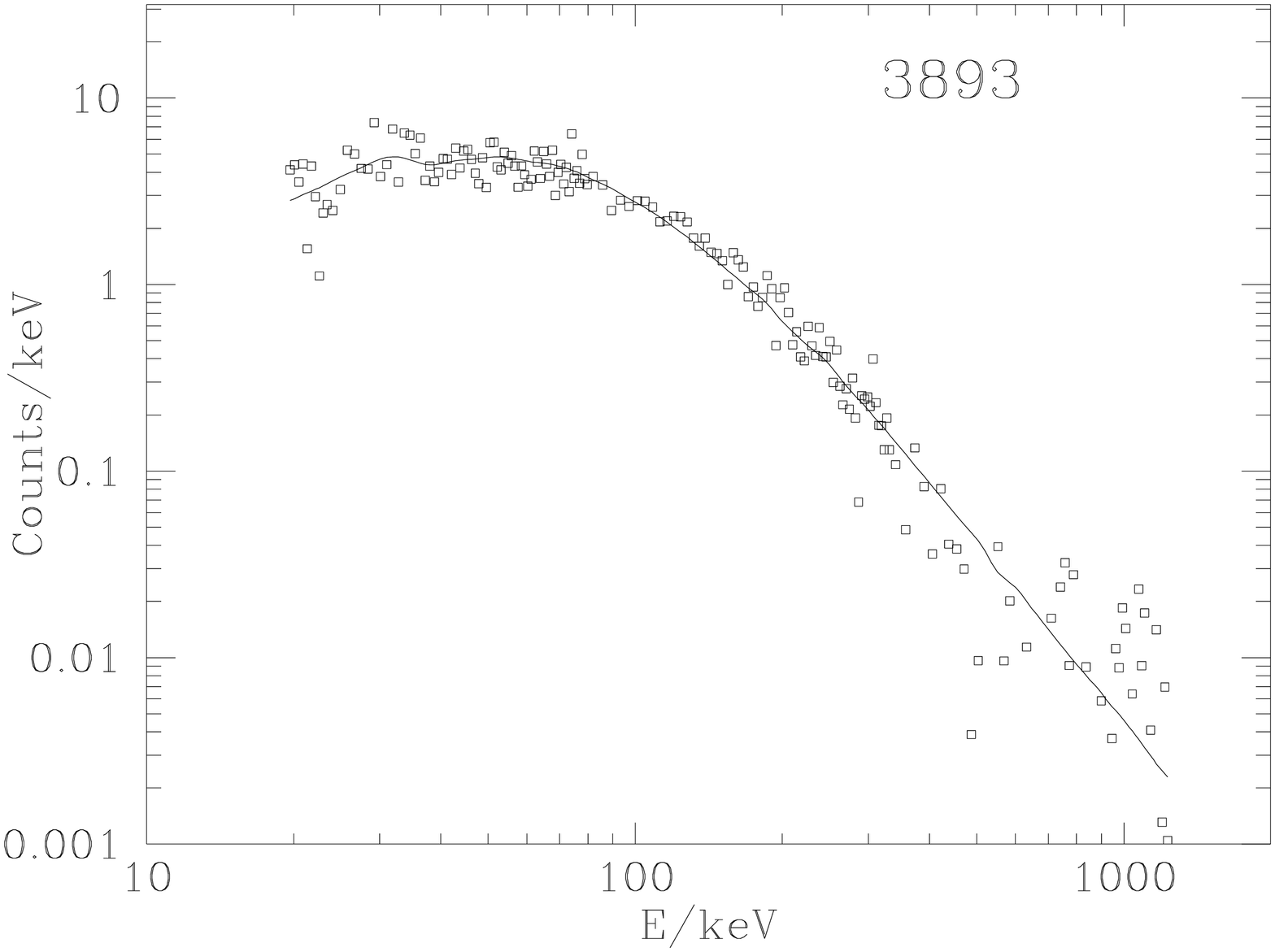,width=.5\textwidth,height=0.5\textwidth,angle=270}}
\caption{Synchrotron fits to photon spectra of BATSE GRBs with trigger
numbers 3253 (left panel) and 3893 (right panel). We display the
count spectrum, which is the photon spectrum, convolved with the
detector response matrix.  An optically thin spectrum with $h\nu_{m} 
= 44$ keV gives
the best fit to 3253, while and optically thick (to synchrotron
self-absorption) spectrum with $h\nu_{m} = 52$ keV and $h\nu_{a}
\approx 40$ keV best fits 3893.}
\end{figure}
Figures 3a and 3b show the spectral fits
for 2 GRBs in our sample (burst triggers 3893 and 3253). 
An optically thin spectrum is the best
fit to 3253, while a self-absorbed
spectrum is required for 3893 - the rollover at low
energies expected from synchrotron self-absorption is
evident, and clearly cannot be accomodated by an
optically thin spectrum.  
[Note that these are {\em counts} spectra and not photon spectra, so
the bumps in the data and the model are a result of the convolution of
the impinging photon spectrum and the model convolved with the detector
response matrix.] 

\subsection{The Electron Energy Spectrum}
 In most models of synchrotron emission, the
 electron distribution is modeled by a power law with
 with spectral index $p$ (as done above).
  Since the high energy index $\beta = -(p+1)/2$ (or
  -(p+2)/2 for a simple cooling model), and
  the majority of bursts have $\beta > -1.5$ (see Figure 1),
  this means that the index $p>2$.  As a result, a 
  cutoff to the power law distribution at some
  minimum energy $\gamma_{m}$ must be imposed to prevent divergence
 of the accelerated electron energy distribution at low energies.
 However, as discussed in the
 introduction, such a sharp cutoff is not a realistic - even stable - scenario.
Instead, one expects
to produce
a gradual 
 cutoff to the electron distribution (perhaps due to such effects
 as the two stream 
instability). In general, particle acceleration in relativistic shocks is not well
understood, and could produce a range of low energy power law
tails (among more complicated behavior) to the electron distribution, or
even a nearly flat low energy electron spectrum.
In order to account for this range of possible low energy
behavior, we
characterize the electron distribution
by the following equation:
\begin{equation}
N(\gamma) = N_{o}\frac{(\gamma/\gamma_{m})^{q}}{1+(\gamma/\gamma_{m})^{p+q}}
\end{equation}
where, now, $\gamma_{m}$ 
is some critical energy below which
the electron distribution changes from
$p$ to $q$ . For $\gamma \gg \gamma_{m}$, $N(\gamma) \propto \gamma^{-p}$, 
while for $\gamma \ll \gamma_{m}$, $N(\gamma) \propto \gamma^{q}$.  Hence,
 $q$ characterizes the ``smoothness'' of the
cutoff (note for $q \rightarrow \infty$,
$\gamma_{m}$ becomes a sharp cutoff to the distribution as defined in the
previous section). 
   
{For this new electron distribution,
an optically thin synchrotron spectrum for an isotropic pitch angle
distribution
takes the form:
\begin{equation}
F_{\nu} = {\cal C}(\nu/\nu_{m})^{(q+1)/2}\int_{0}^{\infty}dx
\frac{x^{-(q+1)/2}}{1+((\nu/\nu_{m})^{(q+p)/2}x^{-(p+q)/2})}
\int_{x}^{\infty} K_{5/3}(z)dz
\end{equation}
where, again, $\nu_{m} = (3/2)\Gamma \gamma_{m}^{2}\nu_{B}$,
 and ${\cal C}$ is the normalization.}
 \begin{figure}[t]
\leavevmode
\centerline{
\psfig{file=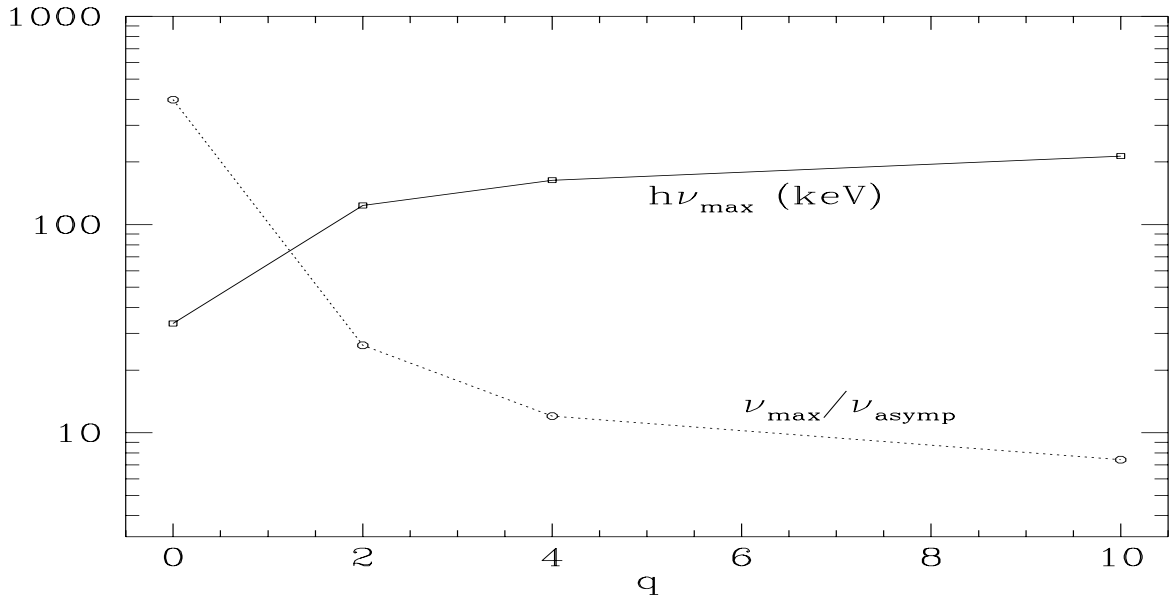,width=1.\textwidth,height=0.5\textwidth,angle=270}}
\caption{The peak of the $F_{\nu}$ spectrum,
$h\nu_{max}$, as a function of $q$ (solid line), and  the
ratio of $h\nu_{max}$ to $h\nu_{asymp}$ (dotted line) where $h\nu_{asymp}$
is defined as the energy in which the optically
thin spectral index is within 5\% 
of its low energy asymptotic value of $1/3$.  Note that the dependence
on $q$ for the latter curve
is quite strong for $q \la 3$, the smoothest cutoffs.
}
\end{figure}
{Note that for $q<-1/3$, the asymptotic photon index
$\alpha = \frac{q+1}{2} - 1 < -2/3$, but for $q>-1/3$, this index 
$\alpha=-2/3$.  This simply means that for $q < -1/3$, the superposition
of the photon spectra from individual
electrons below $\gamma_{m}$ becomes powerful enough to change the low
energy behavior.  That is, there is a ``competition'' between
$\nu^{1/3}$ (the low energy photon spectrum for an individual
electron) and $\nu^{(q+1)/2}$ (the superposed low energy behavior
of the electron distribution below $\gamma_{m}$) -
when $(q+1)/2 < 1/3$, the low energy behavior below $\nu_{m}$ is
different than the usual $\nu^{1/3}$.
Nonetheless, even for $q>-1/3$, the ``smoothness''
of the cutoff can change the spectrum of the emitted photons significantly,
primarily making the transition from the high energy index $\beta$ to
the low energy asymptotic value of $\alpha=-2/3$ more gradual.
Examples of spectra with $q \ne \infty$ are shown in Figure 2 by
the dotted ($q=0$) and short dashed lines ($q=2$).
We point out that as the cutoff of the electron distribution
is made smoother (q smaller),
{the peak of the spectrum shifts to lower energies, while 
the asymptotic value of the spectral index is approached
at a slower rate. 
Figure 4 shows both of these effects.  
The solid line shows the peak of the $F_{\nu}$ spectrum,
denoted as $h\nu_{max}$, as a function of $q$; the dotted line shows the
ratio of $h\nu_{max}$ to $h\nu_{asymp}$, where $h\nu_{asymp}$
is defined as the energy in which the optically
thin spectral index is within 5\% 
of its low energy asymptotic value of $1/3$.

\subsection{Small Pitch Angle Emission}
   The usual analysis of synchrotron radiation assumes electrons are
   distributed isotropically in a either a uniform or
 randomly oriented magnetic field geometry (Pacholczyk, 1970). 
However, this is a simplifying assumption and sometimes
one expects a non-isotropic pitch angle distribution,
usually beamed along the field lines. If the
beaming is strong so that most electrons have
   pitch angles $\theta < 1/\gamma$ (the Lorentz factor of the electron),
   the shape of the synchrotron spectrum changes significantly  
 (see  Epstein, 1973, 
     and Epstein \& Petrosian, 1973).
 An efficient method of producing
 isotropic pitch angle distributions at relativistic energies
 is by plasma turbulence, which can scatter and change the energy
 of the electrons.  For high density, low magnetic field plasmas the Alfv\'en phase 
velocity is less than the speed of light and therefore the speed of the 
particles (relativistic in our case). In this case, the pitch angle diffusion 
rate of the electrons interacting with plasma turbulence is 
much larger than the acceleration rate; consequently, the accelerated electrons
will have an isotropic pitch angle distribution. However, for the low density, 
high magnetic field condition expected for the sources of GRBs the opposite is 
true. In this case the fluctuation in the electric field of the waves 
exceeds the fluctuation of the magnetic field so that the above situation is 
reversed
(see e.g. Dung \& Petrosian, 1994 or Pryadko \& Petrosian 1997). Then the 
pitch angle distribution of the accelerated electrons could become highly 
anisotropic as required in the small pitch angle model.
 One consequence of a small pitch angle distribution
 is that the resulting spectrum at low frequencies is approximately:
 $F_{\nu} \propto \nu$, for $\nu<\nu_{s}$, where $\nu_{s}$
 characterizes the where small pitch angle scattering is
 important. Above this frequency,
  $F_{\nu}$ follows the usual synchrotron
 spectrum (again, we have assumed
 that $\theta \gamma \ll 1$, where $\theta$ is the electron
 pitch angle.  This leads to a single particle emissivity
 that goes as $\nu$ up to $2 \gamma \nu_{B}$, and sharply cuts
 off after that; see Epstein, 1973, \S II).
 { An example of a synchrotron spectrum with a small
 pitch angle distribution in this regime
  is shown by the long dashed line in Figure 2.
 Note that this is an optically thin spectrum with $\alpha=0$; this
 is above the so-called ``line of death'' $\alpha=-2/3$.  
 It should be noted that a very similiar low energy spectrum is 
 obtained when an {\em isotropic} distribution of electrons is
 embedded in a region with a very tangled magnetic field.  If
 there exist magnetic field fluctuations with a
 correlation length that is less than the
 Larmor radius of the electrons, then there will be transverse deflections
 of the electrons with angles less than the relativistic beaming angle.
 Emission due to these fluctuations is very similiar to the small
 pitch angle case and therefore also 
 produces a spectrum $F_{\nu} \propto \nu$ at low energies
 (Medvedev, 2000).  Note that the latter
 scenario imposes restrictions on the structure of the magnetic field,
 while the former (small pitch angle) model constrains the acceleration
 mechanism.  

\section{The Low Energy Asymptotic Behavior} 
  We have shown so far that
synchrotron emission can result in low energy
spectral index values of $\alpha = $ $d$log($F_{\nu})/d$log$(\nu)-1
=-2/3, \ 0, \ , 1.5$ (and possibly $-1.5$ if ``cooling''
effects are important).  The observed distribution
shown in Figure 1, on the other hand, shows a broad
and continuous distribution of $\alpha$ with about
$96$\% of bursts in the range $-1.5<\alpha<0$.  We propose
 that this dispersion is caused by the finite bandwidth of
 the BATSE instruments and arises from the
 fitting of a phenomenological model to the data.
 The theoretical values stated above represent the
 asymptotic logarithmic slopes far from the break
 or peak photon energy.  How well a fitting algorithm
 can determine these asymptotic values depends on 
 whether the asymptotic value of the spectrum is reached
 within the detector window.  This depends both on how ``quickly''
 the spectrum reaches its asymptote and the value of 
 $E_{p}$ relative to the lower end of the detector window.
 In other words, we need to understand
how well spectral fits
can determine the low energy asymptote for the different
cases presented above.  For example,  as 
$E_{p}$ moves to lower and lower energies, we get less and less of
the low energy portion of the spectrum; in this case, our spectral
fits probably will not be able to determine the asymptotic index and will
measure a smaller (softer) value of $\alpha$.  Preece et al. (1998a)
pointed out this effect and attempt to minimize it by
defining an effective low energy index, $\alpha_{eff}$, 
which is the slope of the spectrum
at 25keV (the edge of the BATSE window).  However, a correlation
between $\alpha_{eff}$ and $E_{p}$ will still exist if the
asymptotic value is not reached by 25keV.  This difficulty
becomes more severe the smoother the cutoff to the electron distribution,
because the spectrum takes longer to reach its asymptotic index.

To determine the extent of  these effects, we
simulated data from optically
thin synchrotron models with different values of the
parameters $\nu_{m}$ and $q$ (which
determine the values of $E_{p}$ and $\alpha$ in a Band
spectral fit).  We have included spectra both for an isotropic
pitch angle distribution and for a distribution
of electrons having small pitch angles only.
 Note that since we take an optically thin spectra with
$q \ge 0$, all of  the spectra have a low
energy asymptote of $-2/3$
(isotropic pitch angle distribution)
 or $0$ (small pitch angle distribution).  We normalize the
spectra so that the peak photon flux in the range $50-300$ keV
is $10 \rm ph/cm^{2}/s$ (e.g. a fairly bright GRB).  Our data
points are then drawn from a Poisson distribution with a mean
at the value given by the synchrotron photon spectrum at a particular
frequency; that is, the data points are drawn from
a Poisson distribution with a mean $N_{\nu,i} = (F_{\nu}/h\nu)_{i}$,
where $i$ indexes the data point.  We then
fit a Band spectrum to this data using a conservative
estimate that BATSE is sensitive to all  
photons above $10$ keV.
\begin{figure}[t]
\leavevmode
\centerline{
\psfig{file=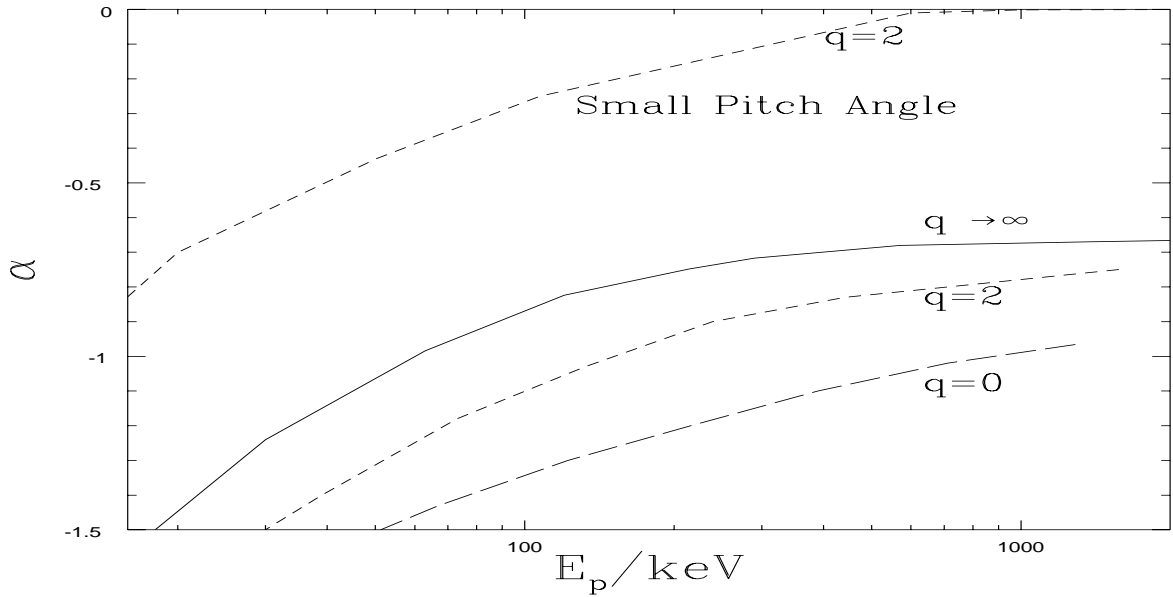,width=1.\textwidth,height=0.5\textwidth,angle=270}}
\caption{The low energy photon index $\alpha$ vs. $E_{p}$, the peak
of $\nu F_{\nu}$, for optically thin spectra with different values
of $q$ for an isotropic pitch angle distribution (lower
three curves), and a small pitch angle
distribution (upper curve).  The values of $q$ (smoothness of
electron spectrum cutoff) are labeled. Note that the smoother the cutoff
the electron energy distribution (lower $q$) the longer it takes the
spectrum to reach the optically thin asymptote within the BATSE window.
}
\end{figure}
Figure 5
shows the value of $\alpha$ as a function of $E_{p}$, for
different degrees of the smoothness of the electron energy distribution 
cutoff in the isotropic pitch angle case (three lower curves),
and for an intermediate cutoff in the small pitch angle distribution
case (upper curve).
Not surprisingly, there is a strong correlation
between the value of $E_{p}$ and the value of
the  index, $\alpha$.  To make sure this isn't purely an artifact
of the Band function, we also tried a broken power law fit ($A(E) \propto
E^{\alpha}$, $E<E_{break}$, $
A(E) \propto E^{\beta}$, $E>E_{break}$, where $A$ is in photons/cm$^{2}$/s/keV) to 
our simulated data and found a very similiar relationship (although
the power law does not give as good of fits as the Band spectrum, so
the relationship was noisier).
Because GRBs have a relatively broad distribution of
$E_{p}$, the above correlation between $\alpha$
and $E_{p}$ will lead to a dispersion in the distribution
of $\alpha$; to determine the extent of this
effect, we need the distribution of $E_{p}$'s.  In our analysis,
we use the distribution observed by BATSE, plotted in Figure 6 (solid
line).  Note that there has been some controversy over the dispersion
in this distribution and we have shown (Lloyd \& Petrosian, 1999) that
indeed this distribution suffers from selection effects in the BATSE
spectral window that tend
narrow the it. These effects are evident in Figure
6, which
also plots $E_{p}$ distributions observed by SMM (sensitive to higher
energies than BATSE; Harris \& Share, 1998) and GINGA 
(sensitive to lower energies than BATSE; Strohmeyer et al., 1998).
\begin{figure}[t]
\leavevmode
\centerline{
\psfig{file=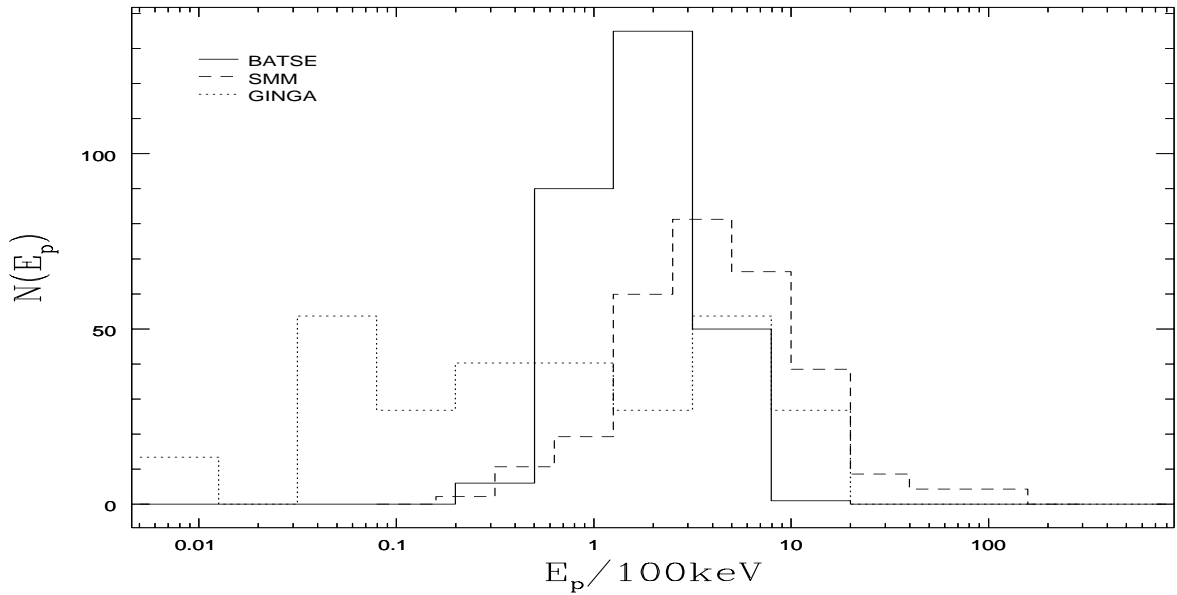,width=1.\textwidth,height=0.5\textwidth,angle=270}}
\caption{
The superposed BATSE, SMM, and GINGA
$E_{p}$ distributions, $N(E_{p}$.}
\end{figure}
The former extends the BATSE $E_{p}$ distribution on the upper end, while
the latter extends it on the lower end.   However, we are interested in
the BATSE {\em observed} $\alpha$ distribution through its relationship
with the BATSE {\em observed} $E_{p}$ distribution.  Hence, the intrinsic
$E_{p}$ distribution (with selection effects accounted for) - although
relevant for understanding other physical aspects of the radiation
processes - is not relevant for our discussion of the observed $\alpha$
distribution.

\section{The Observed $\alpha$ Distribution}
From the observed distribution of $E_{p}$ (Figure 6, solid histogram),
we can determine how the correlation between
$E_{p}$ and $\alpha$ introduced by the fitting procedure
(Figure 5) ``smears'' the distribution of $\alpha$ away from the expected narrow 
distribution around the physical asymptotes of $-2/3$, $0$, etc..
 We represent the correlation
between $\alpha$ and $E_{p}$ by an approximate simple analytical
function; log$(E_{p}) = h(\alpha)$, where the function $h(\alpha)$
depends on the specifics of the synchrotron model (see Figure 5).
{We then approximate the
$E_{p}$ distribution, $f({\rm log}(E_{p}))$ by a Gaussian in
 log$(E_{p})$, 
with a mean and dispersion equal to those of the
observed distribution. 
The distribution of $\alpha$ is then obtained from
the relation 
\begin{equation}
 g(\alpha) = f({\rm log}(E_{p}))\frac{dh(\alpha)}
 {d\alpha}.
\end{equation}
Figure 7 
shows the resultant $\alpha$ distributions in the case of an isotropic
pitch angle distribution for
a sharp ($q=\infty$, right solid curve), intermediate ($q=2$, middle short-dashed
curve),  and flat ($q=0$, left long-dashed curve)
cutoff to the electron energy spectrum, as well as the
small pitch angle distribution case for an intermediate cutoff (right
dot-dashed line). The dotted histogram is the observed
$\alpha$ distribution shown in Figure 1. 
\begin{figure}[t]
\leavevmode
\centerline{
\psfig{file=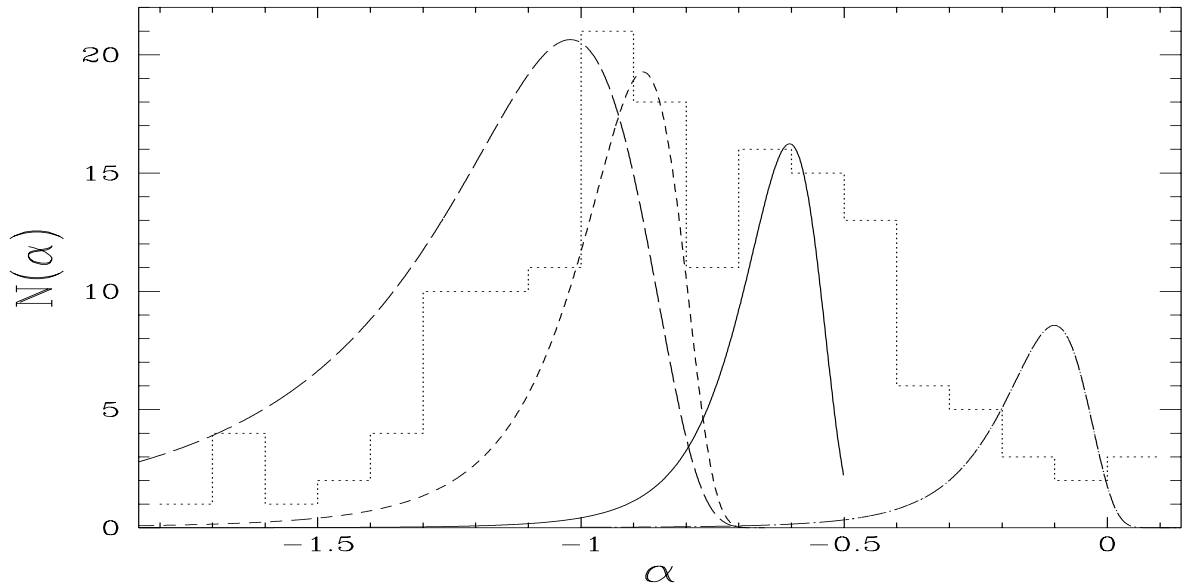,width=1.\textwidth,height=0.5\textwidth,angle=270}}
\caption{The predicted $\alpha$ distributions, $N(\alpha)$, given
the 
 the relationships between $\alpha$ and $E_{p}$ shown in Figure 5.
 The three left curves are for an isotropic pitch
 angle distribution for a $q \rightarrow \infty$
 (right solid curve), $q=2$ (middle short-dashed curve), and $q=0$
 (left long-dashed curve) cutoff to the electron distribution,
  while the rightmost curve is the
small pitch angle distribution case for an intermediate cutoff (dot-dashed
curve).
 The dotted histogram is the observed distribution taken from Preece et
 al. (1999) also shown in Figure 1. The simulated curves
have been separately normalized to the height of this observed
distribution.
}
\end{figure}
 The simulated curves
have been arbitrarily normalized to the height of the observed
distribution at their central values of $\alpha$.  We note
the following:

 1) It appears that almost the entire range of the observed distribution
can be covered by these optically thin synchrotron models.
In particular,
 given a distribution in $q$, an instantaneous 
optically thin spectrum in the large
pitch angle scattering regime 
can easily accomodate bursts below the ``line of death''
with $\alpha \la -2/3$, where most
of the bursts are located.  

2) The second conclusion is that the electron  energy distribution
below the turnover energy $\gamma_{m}$ must be falling off, or must be
(at most) flat ($q \ge 0$).  Otherwise, the optically thin model would predict
too many bursts with $\alpha$ less than about $-1.5$.  

3) We have shown than an optically thin spectrum from a small pitch angle
distribution and a smooth cutoff to the electron energy distribution can
explain all bursts with $-2/3 \la \alpha \la 0$.

4) However, this is not necessarily the only explanation.
Bursts with $\alpha > -2/3$ all the way up to $3/2$ 
can also be due to absorption effects.
  As shown in \S 2, for some bursts synchrotron
self-absorption is necessary to provide a good spectral fit to the data.
However, we usually do not see the values $\alpha=3/2$ ($\nu_{m}<\nu \ll \nu_{a}$)
or $\alpha=1.0$ ($\nu \ll {\rm min}[\nu_{m},\nu_{a}]$) expected
from self-absorption (there are, however, some bursts with such sharp
breaks; see Preece et al., 1999).
If self-absorption is important, $h \nu_{a}$ must be near the lower
edge of the BATSE spectral window.  In this case, we just begin
to see the absorption cutoff and the steep asymptotic low energy
slopes are not yet reached.  Indeed, this is the case for the fits mentioned
in \S 2.

5) The analysis above becomes more complicated when there are two or
more spectral breaks in the detector bandpass (there is some evidence
for two breaks in GRB spectra; see Strohmeyer et al., 1998).  In this case, a fit
to a phenomenological model with a single break will lead to averaging
of the slopes above and below the break it did not fit.  For example,
if $\nu_{m}$ and $\nu_{a}$ are both
present in the spectral window with $\nu_{m}>\nu_{a}$  
and the fit places $E_{p} \propto \nu_{m}$, then
 the Band function will not
accomodate the additional absorption break.  As a result, the low energy
index ends up being a weighted average of the optically thin (-2/3)
and optically thick (1) indices.  If the fit places $E_{p} \propto \nu_{a}$,
then the high energy index $\beta$ will be an average of $-2/3$ and $-(p+1)/2$.
This of course applies to any other two (or more) characteristic frequencies
as well.

6) It should also be noted that we have included only a small amount
of noise (from counting statistics) in our simulations - adding more
noise will contribute to the spread in the simulated $\alpha$ distributions
and strenghten our arguments.

7) Finally, we note that other effects such as inhomogeneities in
the electron distribution (Granot et al., 2000) an other radiative transport
effects (Grusinov \& Meszaros, 2000, Dermer \& Boettcher, 2000)
 can also produce spectra both below
and above the values of $\alpha=-2/3$.

\section{Spectral Evolution}
Not only do emission models have to accomodate the shape of the
observed distributions, but also the temporal behavior of the spectral
parameters.  
The behavior of the spectral characteristics with time throughout 
a GRB can give us
information
about the environment of the local emission region
and conceivably constrain the emission mechanism. 
If the above interpretation of the $\alpha$ distribution is correct, 
then we expect considerable correlation between the variation of 
the observed values of $\alpha$
and $E_{p}$ throughout a burst.
\begin{figure}[t]
\leavevmode
\centerline{
\psfig{file=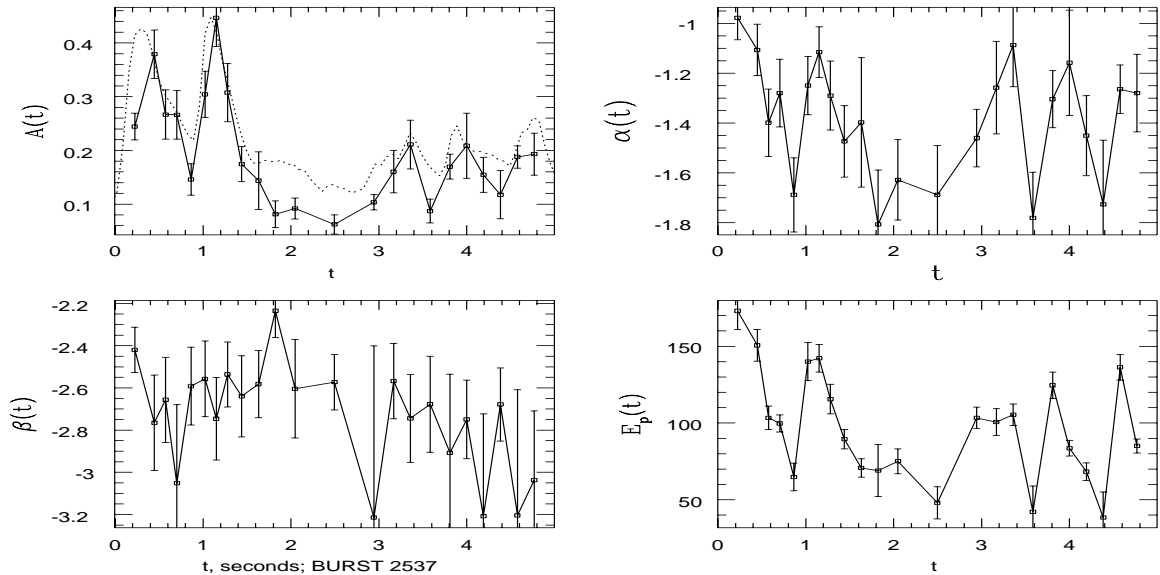,width=1.\textwidth,height=0.5\textwidth,angle=270}}
\caption{Temporal evolution  of the four parameters
characterizing a Band spectral fit - the low energy photon index
$\alpha$ (top right corner), the normalization $A$ in
units of photons/cm$^{2}$/s/keV (top left corner),
the high energy photon index $\beta$ (bottom left corner), the peak
energy $E_{p}$ (bottom right corner) for burst 2537. The time profile of
the GRB with 
with 64ms time resolution (and arbitrary normalization)
is the dotted line superimposed on the
normalization curve.}
\end{figure} 
  Many studies have looked at time evolution
of spectral parameters (e.g. Norris et al., 1986, Ford et al., 1995, Crider
et al., 1997, Preece et al., 1998(b)). 
Ford et al. analyze the evolution of $E_{p}$ for
37 bright, long duration GRBs observed by BATSE.  This study presents evidence
of a general envelope of {``hard-to-soft''} evolution of
$E_{p}$ 
throughout the duration of the burst.  The behavior is explained
in the context of an external shock model in which there is a gradual
decline of average energy as more particles encounter the
shock; the emission mechanism is unspecified.}
{Crider et al. investigate the behavior of the low
energy spectral index $\alpha$ for a sample of 30 BATSE
GRBs.  They find that 18 of these bursts show
{hard-to-soft}
evolution, while 12 exhibit so-called {``tracking''} behavior -
when the evolution of $\alpha$ correlates with the burst
time profile.  All of these bursts also show a strong correlation 
between $\alpha$ and the peak energy, $E_{p}$, as a function of time.
The authors attribute the hard to soft behavior is  to Thomson thinning
of a Comptonizing plasma.  The ``tracking'' behavior is
not explained.}

Recently, Preece et al. (1999)
published a catalog
of spectral data with high time resolution fitting. 
We present the time evolution of three of these
bursts (Figures 8, 9, and 10), plotting all of the
variables that parameterize the (Band) spectral
fits.  Starting in the upper left hand corner and going
clockwise we plot the evolution of the normalization $A(t)$
(in units of photons/cm$^{2}$/s/keV),
$\alpha$, $E_{p}$, and the high energy photon index
$\beta$. [We point out that the time
resolution of the spectral fitting is sometimes coarser than the
real time variation of the burst (at least on the shortest
detector timescale (64ms)).  This will lead to some averaging
effects which may weaken the correlation we expect.  We have
included the higher resolution time profiles (in arbitrary units)
superposed on the plot
of the normalization $A(t)$,
as a reference.]
Given the expected correlation
between $\alpha$ and $E_{p}$ from instrumental and fitting
procedural effects discussed in the previous section, we expect
evolution of $\alpha$ to some extent mimic the evolution of
 $E_{p}$ in time. 

 Indeed this is what we see in Figure 8 -
 the parameter $\alpha$ ``tracks'' $E_{p}$ (and both of these
 track the flux, $A$).  Note the values of $E_{p}$ and
 $\alpha$ are consistent with the case of a flat ($q=0$) 
 cutoff to the electron distribution, but there may be indication
 of variation of $q$ as well.

\begin{figure}[t]
\leavevmode
\centerline{
\psfig{file=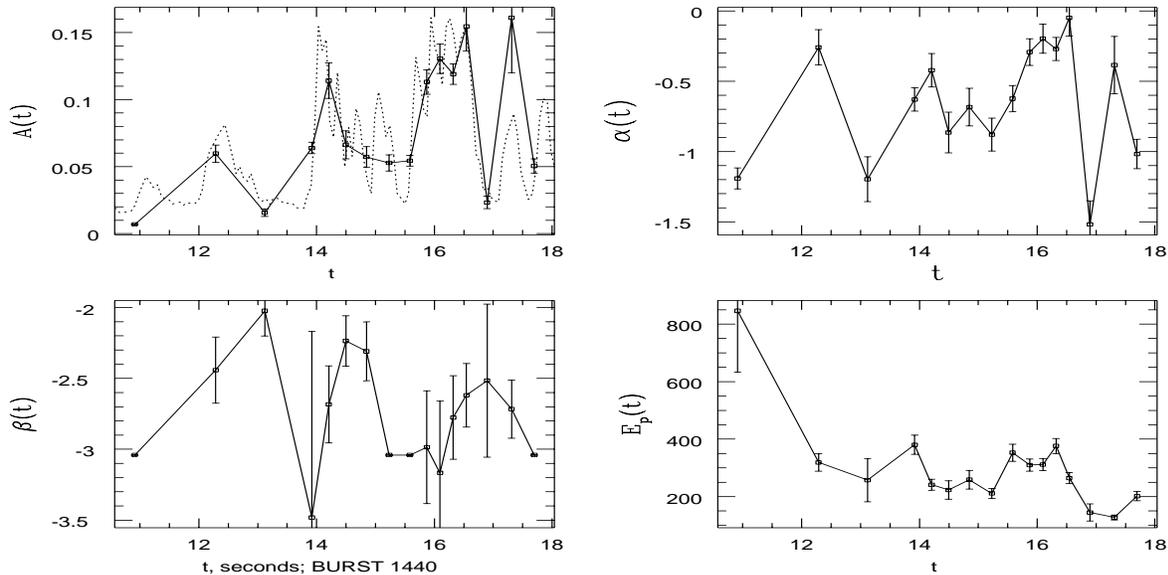,width=1.\textwidth,height=0.5\textwidth,angle=270}}
\caption{Same as Figure 8, but for BATSE trigger number 1440.}
\end{figure}  
  However, it is
 not always this simple.  
In an internal shock scenario, we can regard each pulse in the
time profile as a separate emission episode; in this
case, the internal parameters can vary depending on 
the physical conditions at each shock.  In particular, a change in $q$ 
can create a change in $\alpha$ from
pulse to pulse, independent of
$E_{p}$ (although we
might expect some anti-correlation with $E_{p}$ in this case - high
$q$ gives a high $\alpha$,  but decreases $E_{p}$ if all other paramters 
are the same).  Figure 9 shows such behavior.
The parameter $\alpha$ tracks  the flux;
$E_{p}$ varies on the
same timescale as the flux, but these variations are
superimposed in an envelope of hard-to-soft
evolution. This can be explained by a sharpening of the cutoff
of the electron distribution from peak to peak.  That is, in the first
peak, we have fairly high $E_{p}$ values ($\sim 500$ keV) and moderately
low ($\sim -1$) values of $\alpha$.  This is what we expect for a $q \sim 0$ or $1$
cutoff.  In the second peak, $E_{p}$ is about $300$ keV, but the
$\alpha$ value is around $-0.6$ - this is what we'd expect for a sharper cutoff
($q \ga 5$).

In addition, we may see evidence of evolution of the
opacity or pitch angle distribution from pulse to pulse.
For example, in Figure 10, the parameter $\alpha$ appears to evolve from hard (above
the ``line of death'') to soft, while $E_{p}$ appears to tracks the normalization.
In the early phase, $\alpha > -2/3$ and $E_{p}$ is moderate ($\sim 300$ keV).
At later times, $E_{p}$ is at fairly high values ($\ga 500$ keV),
while $\alpha$ is fluctuating around the
expected asymptote for optically thin emission, $\alpha = -2/3$
(this is what we expect from the correlation discussed in this paper -
when
$E_{p}$ is high enough, the asymptote of the spectrum is reached
inside the BATSE window).
The overall behavior can be attributed either to a transition from a small
pitch angle to isotropic regime, or to a transition
from an optically thick (to synchrotron self-absorption) to optically
thin regime.  
Note that we have not accounted for any biases determination
of the high energy photon index $\beta$ might cause. 
 This ultimately can affect the value
of $\alpha$ or $E_{p}$.  For most cases, $\beta$ remains
fairly constant, with only minor fluctuations.
\begin{figure}[t]
\leavevmode
\centerline{
\psfig{file=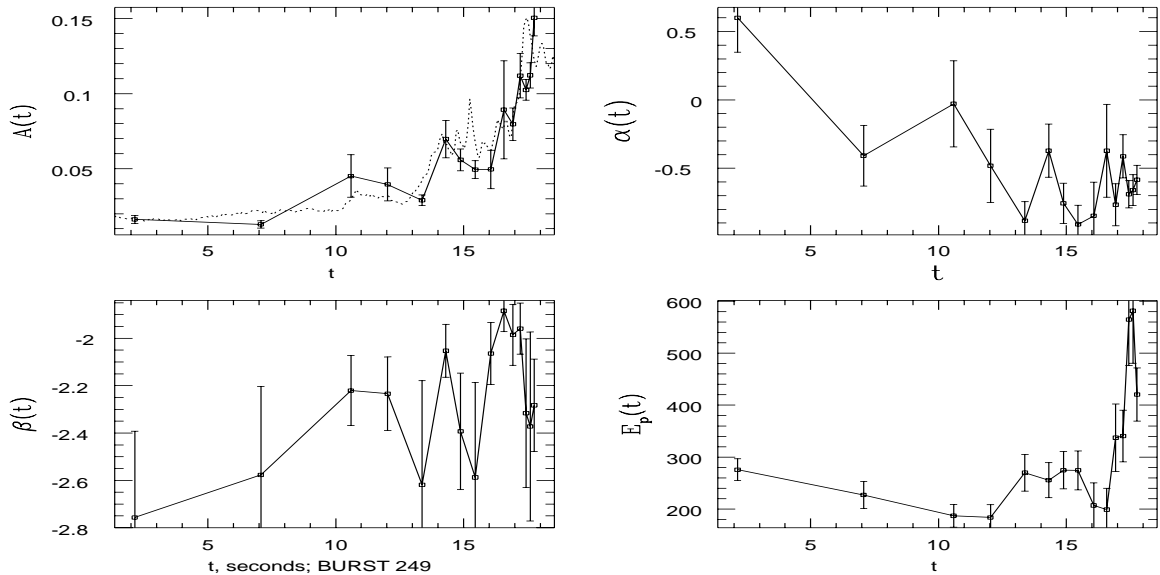,width=1.\textwidth,height=0.5\textwidth,angle=270}}
\caption{Same as Figure 8, but for BATSE trigger number 1440.}
\end{figure}

Our purpose here is not to present a rigorous temporal analysis of
the spectral parameters, but to show the diversity of spectral evolution
from burst to burst,
and to discuss both the procedural (i.e. spectral fitting) and physical
reasons that produce this behavior.
The variety of observed
spectral evolution and its meaning in terms of synchrotron
emission from internal shocks will be discussed in much more detail
in an upcoming publication.

%
%
%

\section{Summary and Conclusions}
  Although synchrotron radiation has been suggested to
  be a viable 
  emission process for GRBs (Katz, 1994, 
  Tavani, 1996), there has
  since been much
  controversy over what radiation mechanism actually gives rise
  to the observed GRB spectra.
  Several studies (e.g., Ghissilini et al., 1999, Celotti \& Ghissilini, 1999)
   have attempted to rule
out synchrotron radiation as the source of GRB spectra based on certain
discrepancies between the model's predictions for the
low energy spectral behavior and the observed
data.  In particular, the simple model of 
synchrotron radiation in an optically thin medium 
from a power law distribution of
electrons with an isotropic pitch angle distribution 
 (and neglecting radiation losses)
predicts a value of the
low energy photon index $\alpha$ of $-2/3$.   The observed distribution of
$\alpha$ is clearly inconsistent with this predicted behavior.
The predicted value of $\alpha$ expected in a 
a simple cooling spectrum  is
$-3/2$, which is also inconsistent with the data.  Furthermore
it has been suggested (Crider et al., 1997)
that the {\em variation} of $\alpha$ throughout
the burst's duration is inconsistent with what is expected from these
synchrotron models.
 
 In this paper, we show that synchrotron radiation can in fact explain
the observed spectral behavior of GRBs.  We show that the source of
the discrepancies described above are a result of two factors:
a) the simplistic assumptions
made in standard synchrotron models, b) assuming
that the spectral fitting procedure accurately determines
the asymptotic values of the spectral indices, without accounting for the effects
of the detector bandwidth in the fitting procedure. [The latter
effect must be accounted for not only when comparing
to synchrotron models, but all
other emission models as well.]

We focus on the so-called ``instantaneous'' synchrotron spectra (in which
radiation losses are not evident).
As mentioned above,
the simple cooling spectrum 
is not consistent with observations.
A cooling spectrum
is obtained when electrons are injected in a magnetized region, 
where they then lose all of their energy to radiation; it is not clear
that this is a viable scenario for GRBs.  Athough the ``cooling time'' 
(the timescale over which electrons lose their energy to radiation) is much
shorter than other timescales involved in the fireball model (e.g. the hydrodynamical
timescale), it is probably unrealistic to separate the acceleration and
radiation loss processes at the site of the GRB.  To sustain the burst,
there should be continual acceleration of the emitting particles
with a rate equal or greater
than the synchrotron loss rate.  In this case, particles can still lose
all of their energy to radiation quickly, but we will have the 
instantaneous 
spectrum 
as a result of the competing acceleration and loss processes. [Note
that because electrons are losing all of their energy to radiation,
the efficiency (i.e. the ratio of energy radiated to the total
energy of the GRB) in this model is the same as calculated using
the simple cooling model - about $1$\% (see, e.g., Kumar, 1999).]
We believe
that this is an attainable scenario, although proof of this
is beyond the aim of this paper. 

  To investigate whether the instantaneous synchrotron models
  are consistent with the existent data, we
relax four important assumptions used in the usual analysis with 
synchrotron models: 1) We allow  for the possibility 
of synchrotron self-absorption, which produces a hard low energy
photon index of either $1$ or $3/2$, depending on the
relative values of the synchrotron self-absorption frequency
and the minimum electron frequency. 2) We allow for
a more realistic smooth (rather than sharp) cutoff to
the electron distribution, which tends to soften the low
energy spectral behavior. 3) We include effects of small pitch angle scattering,
which leads to a value of $\alpha$ of $0$. 4)  We account for the fact that as
the break energy of the GRB spectrum approaches the lower edge of the
BATSE spectral window, the low energy spectral index will become softer (the
expected asymptote of $-2/3$ is not reached).
Items 1-3 above produce physically different low energy spectral
behavior.  Item 4 is a procedural and instrumental effect which
leads to additional dispersion of the $\alpha$ distribution.

We show that including all of these effects can explain the observed
distribution of  $\alpha$'s.
We also infer from our analysis that the electron energy distribution must
flatten or decline at low energies; otherwise, we would see many more
bursts with $\alpha<-3/2$.  GRBs whose $\alpha$ values
lie above the ``line of death'' ($\alpha=-2/3$)
 may be explained either by emission from electrons with 
small pitch angles or absorption processes (e.g. synchrotron self
absorption).
 Finally, we present some examples of the temporal behavior of
 GRB spectral parameters.  We show that some
 of the temporal evolution is explained by the expected
 correlation between $\alpha$ and $E_{p}$, that results from
 the detector bandwidth and the fitting process.  Other temporal
 variations in the spectral parameters suggest changes in the
 physical conditions at the GRB source from pulse to
 pulse (where each pulse may represent independent emission episodes, for
 example, in an internal shocks model).  A detailed study
 of the variety of spectral evolution and its consistency with
 synchrotron emission from internal shocks is the subject of an
 upcoming publication.
 
 In this paper, we have dealt only with the low energy spectral
 index $\alpha$.  Similiar discrepancies and limitations exist
 for the high energy spectral index $\beta$.  However, the value of this index is not
 as well determined or reliable because the signal-to-noise
 decreases rapidly with energy, to an extent that it is questionable whether
 we even have a simple power law spectrum above $E_{p}$.  However, there
 is some reliable data for bright bursts, and we shall explore the effects of 
 modifications described here for $\beta$ in a future publication.

Acknowledgements:  This work was supported by the CGRO Guest
Investigator Program and by the Stanford McMicking Fellowship.
We would like to thank Rob Preece for many useful
discussions, as well as providing and helping with the some
of the BATSE data used in this analysis.  We would also like
to thank Pawan Kumar for helpful
comments. Finally, we are greatly indebted to the referee for an extremely
careful reading of the manuscript, and helpful comments which much improved
this paper.


\end{document}